\documentclass[pre,twocolumn,showpacs,preprintnumbers,reprint,amsmath,
                amssymb,superscriptaddress]{revtex4}
\usepackage[latin1]{inputenc}

\usepackage{graphicx} 
\usepackage{dcolumn}  
\usepackage{bm}       
\usepackage{epsfig}

\begin{document}

\newcommand{\be}{\begin{equation}}
\newcommand{\ee}{\end{equation}}
\newcommand{\bea}{\begin{eqnarray}}
\newcommand{\eea}{  \end{eqnarray}}
\newcommand{\bit}{\begin{itemize}}
\newcommand{\eit}{  \end{itemize}}

\title{Generalized Lyapunov exponents of
       the random harmonic oscillator: 
       cumulant expansion approach }

\author{Ra\'ul O. Vallejos}
\email{vallejos@cbpf.br}
\homepage{http://www.cbpf.br/~vallejos}
\affiliation{Centro Brasileiro de Pesquisas F\'{\i}sicas (CBPF), 
             Rua Dr.~Xavier Sigaud 150, 
             22290-180 Rio de Janeiro, Brazil}
             
             \author{Celia Anteneodo}
\email{celia@fis.puc-rio.br}
\affiliation{Department of Physics, PUC-Rio and 
National Institute of Science and Technology for Complex Systems, 
Rua Marqu\^es de S\~ao Vicente 225, G\'avea, CEP 22453-900 RJ, 
Rio de Janeiro, Brazil}

\date{\today}              
              
\pacs{05.45.-a, 05.40.-a, 02.50.Ey}


\begin{abstract}
The cumulant expansion is used to estimate generalized Lyapunov 
exponents of the random-frequency harmonic oscillator.
Three stochastic processes are considered:
Gaussian white noise, Ornstein-Uhlenbeck, and Poisson shot noise. 
In some cases, nontrivial numerical difficulties arise. 
These are mostly solved by implementing an appropriate
importance-sampling Montecarlo scheme.
We analyze the relation between random-frequency oscillators 
and many-particle systems with pairwise interactions like
the Lennard-Jones gas. 
\end{abstract}
 
\maketitle

\section{Introduction}
\label{sec1}

Lyapunov exponents quantify sensitivity to initial 
conditions in dynamical systems.
The existence of a positive Lyapunov exponent implies
that trajectories initially close in phase space 
will typically diverge exponentially fast in time. 
In practice, this sets a limit for predicting the
future behavior of the system, because 
small imprecisions in the knowledge of the initial state 
will be amplified at a rate given by the largest 
Lyapunov exponent.
Even if determinism subsists on a short time scale, on 
longer time windows the system exhibits features of
randomness.
This property lies at the basis of the statistical 
description of many-particle deterministic systems.
Hence the interest in analytical estimates of
Lyapunov exponents in simple statistical-mechanics 
models.

The theory of Lyapunov exponents of hard-ball systems 
has a long history. 
It started with the pioneering work of Krylov 
\cite{krylov,ma}, 
was rigorously developed by Sinai \cite{sinai} and
collaborators, and
completed (to some extent) by van Beijeren, Dorfman and 
co-workers \cite{vanb1,vanb2,vanzon,kruis,dorfman}. 
The analytical calculation of, e.g., the largest 
Lyapunov exponent of a dilute rigid-sphere gas, is 
based on the fact that the dynamics consists of  
free rectilinear motions interrupted by instantaneous 
elastic collisions \cite{vanzon};
the expressions so-obtained agree quantitatively
with the numerical experiments 
\cite{dellago1,dellago2,vanzon}. 

The case of a dilute gas with finite-range interactions 
can be handled in close analogy with the rigid-sphere 
problem: 
though the collisions are not trivial any more, the
dynamics is still ruled by occasional pairwise 
encounters \cite{vanzon,kimball,elyutin}. 
However, when one considers long-range interactions
(or short-range interactions and high densities),
the theoretical approach must be substantially
modified. 

In the general case we must deal with the full system of 
coupled differential equations that govern the evolution 
of multidimensional tangent vectors
$\eta(t)$. 
Consider for instance a gas of $N$ particles in three
dimensions described by the Hamiltonian
\begin{equation}
\label{ham}
{\cal H} = 
\sum_{i=1}^{3N} \frac{ p^2_i}{2m} + {\cal V}(q_1,\ldots,q_{3N}),
\end{equation}
where $q_i$ and $p_i$, are conjugate position-momentum 
coordinates. 
Assuming $m=1$, tangent vectors evolve according to
\begin{equation}
\label{tangent}
\dot \eta = 
\left( 
\begin{matrix}
    0          &    1   \cr
 -{\bf V}(t)   &    0        
\end{matrix}
 \right)\, \eta  \; 
\end{equation}
(dot meaning time derivative), 
where $\bf V$ is the Hessian matrix of the potential ${\cal V}$, 
namely
\begin{equation} 
\label{V}
V_{ij} =  
\frac{\partial^2{\cal V}}{\partial q_i \partial q_j} \;.
\end{equation}
The Hessian depends explicitly on time because it is calculated
along a reference trajectory $q(t)$.
Once initial conditions $z_0=(q_0,p_0)$ and $\eta_0$ have been 
specified,
one can find $\eta(t)$ from Eq.~(\ref{tangent}).
Then the Lyapunov exponent $\lambda$ is obtained by calculating 
the limit \cite{benettin}
\begin{equation}
\label{liap}
\lambda = 
\lim_{t \to \infty} \lambda \left( t; z_0,\eta_0 \right) \;,
\end{equation}
where
\begin{equation}
\label{liapfinite}
\lambda \left( t; z_0,\eta_0 \right) = 
 \frac{1}{t} \ln | \eta (t; z_0,\eta_0)| \;.
\end{equation}
The finite-time Lyapunov exponent $\lambda \left( t; z_0,\eta_0 \right)$
depends on the initial conditions $z_0$ and $\eta_0$.
However, assuming ergodicity on the energy-shell,
$\lambda$ becomes independent of $z_0$, which can then be chosen 
randomly, e.g., according to the microcanonical distribution. 
There will also be no dependence on initial tangent vectors,
because if $\eta_0$ is also chosen randomly, it will always have 
a non-zero component along the most expanding direction.
In spite of being redundant, the averaging over $z_0$ and $\eta_0$
permits to treat equations (\ref{tangent}) formally as a system of 
stochastic differential equations 
\cite{vankampen}.
So, in this ``stochastic'' approach one attempts the 
analytical estimation of the average
\begin{equation}
\label{liapav}
\lambda = 
\lim_{t \to \infty} 
\frac{1}{t} 
\langle 
\ln | \eta (t; z_0,\eta_0)| 
\rangle \;.
\end{equation}
This is a hard task, however. 
It is much simpler to evaluate the 
generalized Lyapunov exponent \cite{benzi,castiglione}
\begin{equation}
\label{liap2}
\lambda_2 = 
\lim_{t \to \infty} 
\frac{1}{2t} 
\ln \langle 
 | \eta (t; z_0,\eta_0)|^2 
\rangle \, ,
\end{equation}
and assume it approximately coincides with the standard
Lyapunov exponent, which is justified in the absence
of intermittency \cite{castiglione}.

Moreover, if the Hamiltonian of the system can be decomposed
as some ``free'' part plus weak interactions, then perturbative 
techniques, like the cumulant expansion \cite{vankampen,kubo,fox}, 
can be invoked.
This is essentially the approach followed by 
Barnett et al. \cite{barnett1,barnett2,barnett3}, 
Pettini et al. \cite{pettini1,pettini2,pettini3},
and the present authors \cite{av1,av2,av3}.
Though there are some differences among the 
formulations of the three groups above, 
it may be said that the main theoretical conclusion 
extracted from that body of work is the following.
As far as $\lambda_2$ is concerned,
if one combines the cumulant expansion with some kind 
of isotropy approximation (which may be fully justified 
in some cases), the original problem of $6N$ differential 
equations can be reduced to a system of only two equations 
for a ``representative" single degree of freedom:
\begin{equation}
\label{tangent2}
\left( \begin{matrix}
    \dot \eta_1   \cr
    \dot \eta_2           
\end{matrix} \right)
 = 
\left( \begin{matrix}
    0          &    1   \cr
 -\kappa(t)    &    0        
\end{matrix} \right)  
\left( \begin{matrix}
     \eta_1   \cr
     \eta_2           
\end{matrix} \right)\; .
\end{equation}
In this kind of mean-field approximation, the ``curvature" 
$\kappa(t)$ is a scalar stochastic  process, whose cumulants 
can be related to the {\em operator cumulants} of the Hessian 
${\bf V}(t)$ (see, e.g., \cite{av1}).
 
The comparison of theoretical results obtained with the
cumulant expansion --truncated at the second order-- 
versus numerical simulations has met mixed 
success.
The agreement is very good for some many-particle systems with
bounded weak interactions \cite{av2,av3} and for the 
Fermi-Pasta-Ulam system \cite{pettini3}. 
On the other side, the results for the 1d-XY model \cite{pettini3} 
and for a dense one-component plasma \cite{barnett1,comment} are 
not so satisfactory.

Anyway, the mentioned tests, which compare theoretical estimates
for $\lambda_2$ against numerical calculations for $\lambda$, 
should be taken with some reservations:
(a) Pettini et al. did not check if the approximate
equality $\lambda \approx \lambda_2$ indeed holds 
\cite{pettini1,pettini2,pettini3}. 
Moreover, their theory includes a fitting parameter 
[the correlation time of the process $\kappa(t)$].
Then, it may happen that the theory really agrees with the 
simulations, or, alternatively, it may be the case of a 
disagreement that is compensated by a suitable choice of the 
correlation time.
(b) The authors of Refs.~\cite{av1,av2,av3} derived 
Eq.~(\ref{tangent2}) from first principles (no fitting
parameters) and verified numerically that 
$\lambda \approx \lambda_2$ holds in their tests.
However, they used a simple (``brute force'' \cite{vanneste10})
Montecarlo sampling for doing the average (\ref{liap2}).
And it is known (e.g., \cite{vanneste10}) that simple samplings 
tend to produce wrong estimates of generalized Lyapunov 
exponents
\begin{equation}
\label{liapq}
\lambda_Q = 
\lim_{t \to \infty} 
\frac{1}{Qt} 
\ln \langle 
 | \eta (t; z_0,\eta_0)|^Q
\rangle \, .
\end{equation}
The larger the value of $Q$, the stronger this spurious effect.
(Consistently, there are no difficulties in the numerical
calculation of the standard $\lambda$, given that 
$\lambda_Q \to \lambda$ for $Q \to 0$.)

In conclusion: if one wants to assess the quality of  
theoretical predictions unabiguously, then it is necessary 
to develop trustable Montecarlo algorithms for the 
calculation of $\lambda_Q$.
We are not aware of the existence of such methods for
Hamiltonian many-particle systems.
On the other side, an importance-sampling \cite{binder} 
algorithm was recently proposed by Vanneste for
calculating $\lambda_Q$ in stochastic dynamical systems. 
The algorithm was shown to perform efficiently for
white noise and $Q$ not too large \cite{vanneste10}.
 
The present work is part of a larger project that aims at
defining the limits of validity of the cumulant 
approach for the Lyapunov exponent of many-particle 
Hamiltonian systems.
We start our investigations with the simplified mean-field 
setting (\ref{tangent2}).
This is the simplest possible case having the same formal
structure as the many-body problem.
By choosing $\kappa(t)$ to be a stochastic process we
shall be able to use importance-sampling in the numerical
calculations.
For several choices of $\kappa(t)$ we shall both analyze 
the performance of the cumulant expansion and test the 
numerical algorithms.

It has been argued \cite{pettini3} that, for typical chaotic
many-body systems, $\kappa(t)$ should be close to Gaussian
white noise; this is the first
case we shall consider.
For Gaussian white-noise the second-order cumulant expansion 
for $\lambda_2$ is exact, thus this case is ideally 
suited for analyzing the difficulties that appear 
in the numerical calculation of $\lambda_2$ (Sect.~\ref{sec4}).

Next, we keep the Gaussian and Markov properties but 
allow for finite correlation times, leading to the 
Ornstein-Uhlenbeck process.
In this case we calculate the fourth cumulant 
contribution to $\lambda_2$.
This test will give us some idea of (i) the convergence 
rate of the cumulant expansion, and (ii) the performance of
the importance-sampling method for colored noise (Sect.~\ref{sec5}).

Last we study the situation of $\kappa(t)$ being Poisson 
white shot-noise. This appears to be the appropriate choice 
for modeling the tangent-vector dynamics in dilute 
gases with short-range interactions. 
Like in the case of Gaussian white noise, here we have 
analytical expressions for the generalized exponents
$\lambda_2$, $\lambda_4$, $\lambda_6$, etc. 
So, this case will provide an opportunity for further testing 
of the numerical algorithm. 
At the same time it will be helpful for characterizing the 
distribution of finite-time Lyapunov exponents, e.g., 
when is this distribution approximately Gaussian? (Sect.~\ref{sec6}).

Section~\ref{sec2} contains a short review of the cumulant expansion
as applied to the determination of some generalized Lyapunov exponents.
In Sect.~\ref{sec3} we describe the three Montecarlo methods
considered in this paper: simple, simple-Gaussian, and 
importance-sampling.
Section~\ref{sec7} presents a summary of our results and the
final remarks.

Before proceeding to the bulk of the paper let us comment
that the random oscillator of Eq.~(\ref{tangent2}) is 
formally equivalent to the Schrödinger equation for
a particle in a disordered potential (Anderson localization 
problem in one dimension).
Thus many useful results related to random oscillators
can be found in the condensed-matter literature 
\cite{paladin87,tessieri00,tessieri01,tessieri02,gurevich09,
lugan09,schomerus02,iomin09,gurevich11}.

\section{Cumulant expansion for the Kubo oscillator}
\label{sec2}

Equation~(\ref{tangent2}) describes a harmonic 
oscillator with a random frequency $\omega$ such that
$\omega^2=\kappa$ (Kubo oscillator). 
It is worth extending this model a bit to account for 
the possibility of damping, i.e., we shall consider an 
oscillator described by the first-order equations 
\bea
\dot q & = & p                             \, , \nonumber \\
\dot p + \alpha \, p + \kappa \, q & = & 0 \, .
\label{kuboosc}
\eea
Let us make the identifications 
$q=\eta_1$, 
$p=\eta_2$ 
\footnote{The equations of motion of the random harmonic 
oscillator being linear, phase space and tangent space
can be identified.
Accordingly the Lyapunov exponent equals one half 
the average rate of energy growth.}.
Then, putting $\alpha=0$ we recover 
(\ref{tangent2}).

Some analytical results for the Lyapunov exponent of the
Kubo oscillator (\ref{kuboosc}) can be found in the 
literature (see, e.g., \cite{mallick,leprovost,peleg}). 
Here we shall concentrate on the generalized exponent 
$\lambda_2$. 
For this purpose we must consider the dynamics of second 
order products:
\begin{equation}  \label{v2dot}
\frac{d}{dt}
\left(
\begin{array}{c}
q^2 \cr
p^2 \cr
qp
\end{array}
\right)
=
\left(
\begin{array}{ccc}
           0 &  0         & 2          \cr
           0 & -2\alpha   & -2\kappa   \cr
     -\kappa &  1         & -\alpha  
\end{array}
\right)
\left(
\begin{array}{c}
q^2 \cr
p^2 \cr
qp
\end{array}
\right) 
\equiv {\bf B}(t)
\left(
\begin{array}{c}
q^2 \cr
p^2 \cr
qp
\end{array}
\right) 
\, .
\end{equation}
Let us assume that both parameters $\alpha$ and
$\kappa$ are stationary stochastic processes.
If fluctuations are small enough (in a sense that will be 
discussed later), one can obtain dynamical equations for 
the second-order {\em averages} using the cumulant 
expansion \cite{vankampen}. 
Splitting the stochastic matrix as an average plus
fluctuations:
\begin{equation} 
{\bf B}(t)={\bf B_0}+{\bf B_1}(t)  \, ,
\end{equation}
it can be shown that for long times one has \cite{vankampen}:
\begin{equation}
\frac{d}{dt}
\left \langle
\left(
\begin{array}{c}
q^2 \cr
p^2 \cr
qp
\end{array}
\right)
\right \rangle
={\bf K}
\left \langle
\left(
\begin{array}{c}
q^2 \cr
p^2 \cr
qp
\end{array}
\right) 
\right \rangle \, ,
\label{eqK}
\end{equation}
where ${\bf K}$ is the $3 \times 3$ matrix given by the 
operator cumulant expansion \cite{vankampen}
\begin{equation} 
\mathbf{K}= 
\mathbf{B}_0 + 
\int_0^\infty 
\left \langle 
\mathbf{B}_1(\tau) 
\, e^{\mathbf{B}_0 \tau} \, 
\mathbf{B}_1(0) 
\right \rangle 
e^{-\mathbf{B}_0 \tau}
d\tau + \ldots \, .
\label{cum}
\end{equation}
Ellipsis stand for third and higher cumulants (some explicit
expressions can be found in \cite{fox}). 
The exponent $\lambda_2$ is related to the eigenvalue
of ${\bf K}$ that has the largest real part: 
\begin{equation} \label{lambdastar}
\lambda_2=
\frac{1}{2} 
\max \,
\Re \, 
\left\{ k_1,k_2,k_3 \right\} \, ,
\end{equation}
with $k_i$ the eigenvalues of $\mathbf{K}$.

Starting from the evolution equations for higher order
products [analogous to (\ref{v2dot})] and repeating the
same steps above, one can derive the corresponding 
expressions for $\lambda_4$, $\lambda_6$, etc.
Of course, the algebraic difficulties increase with the order of
the exponent.

\section{Numerical methods}
\label{sec3}

The numerical evolution of Eqs.~(\ref{kuboosc}) was 
performed by means of the Euler algorithm with time 
step $dt=10^{-3}$ (some higher-order algorithms 
\cite{kloeden} were tested, but did not lead to 
substantial improvements). 
A set of trajectories is generated by randomly choosing 
$(q_0,p_0)$,  $\alpha(t)$ and $\kappa(t)$.
For each trajectory we computed the norm 
$|\eta(t)|=\sqrt{q^2+p^2}$ as a function of time. 
The Lyapunov exponent is then approximated by the average
of finite-time exponents:
\begin{equation}
\label{lambdanum}
\lambda 
\approx
\langle 
\lambda(t; \zeta_0)
\rangle 
= 
\langle 
\frac{1}{t}
\ln | \eta (t; \zeta_0)| 
\rangle 
\;.
\end{equation}
The operation 
$\langle \cdots \rangle$
means averaging over a certain number of realizations of 
the pseudorandom variables (compactly denoted by $\zeta_0$) 
that determine the trajectories. 
Time $t$ must be large enough to assure the convergence of 
the average to the desired precision.

In principle we could use the same scheme as before for 
estimating generalized exponents, i.e.,
\begin{equation}
\label{lambdaQnum}
\lambda_Q \approx 
\frac{1}{Q t} 
\ln 
\langle | \eta (t; \zeta_0)|^Q  \rangle 
=
\frac{1}{Q t} 
\ln
\langle 
e^{Q  t  \lambda(t; \zeta_0)}
\rangle 
\;,
\end{equation}
the last equality following from (\ref{lambdanum}).
However such a simple averaging tends to undestimate 
rare events.
Hence, spurious results are expected whenever the 
distribution $P(\lambda_t)$ does not decay fast enough
\cite{vanneste10,anteneodo10,gurevich11}. 
A somewhat better alternative is, 
instead of straightforward averaging,
to estimate the generalized exponent from
the first terms of the series
\begin{equation} 
\label{local}
\lambda_Q \approx
\sum_{n\ge 1} \frac{(Qt)^{n-1}}{n!} \, \kappa_n(t)\,,
\end{equation}
where $\kappa_n$ are the $n$th-order cumulants of 
$P \left( \lambda_t \right)$ \cite{zillmer03}. 
In principle, these cumulants could be estimated 
numerically.
However, for the samples we considered, third and higher 
cumulants are typically rather unstable \cite{anteneodo10}. 
So, it is practically impossible to assess the convergence 
of the expansion (\ref{local}). 
For this reason, cumulants $\kappa_n$, with $n \ge 3$, 
will not be included in our calculations.
Thus we arrive at 
\begin{equation} 
\label{gaussianap}
\lambda_Q \approx \lambda  + \frac{1}{2} \, Q \, t \, \kappa_2(t)  \, . 
\end{equation}
[If $P(\lambda_t)$ is Gaussian, this expression is exact.]
We call the procedure leading to Eq.~(\ref{gaussianap})
{\em simple Gaussian averaging}.
From Eq.~(\ref{gaussianap}) one can derive approximate expressions 
for the standard Lyapunov exponent, the simplest one being
\begin{equation} 
\lambda \approx 2 \lambda_2 - \lambda_4   \, . 
\end{equation}
Conversely, when $\lambda,\lambda_2,\lambda_4$ are known, 
the deviation from equality in the formula above 
provides a measure of the non-Gaussianity of $P(\lambda_t)$.  

When the tail of $P(\lambda_t)$ is essential for the 
determination of $\lambda_Q$ and it is not Gaussian, 
the approximations 
(\ref{lambdaQnum}) and (\ref{gaussianap}) 
are bound to fail.
In this case one must resort to numerical methods capable 
of sampling the relevant part of the distribution 
$P(\lambda_t)$.
The {\em importance-sampling} Monte Carlo algorithm recently 
proposed by Vanneste is especially suited for our 
needs.
The algorithm, both efficient and easy to implement, uses a 
simple random resampling step: those trajectories which 
contribute the most (least) to the average are cloned (pruned) 
with a large probability \cite{vanneste10}. 

Having presented the theory and the numerical methods,
we are ready to proceed with the comparisons.

\section{Gaussian white noise}
\label{sec4}

Only when the matrix stochastic process $\bf B_1$ is
Gaussian and delta-correlated the cumulant expansion 
does stop at the second order, i.e., Eq.~(\ref{cum}) 
without the ellipsis becomes exact \cite{fox}. 
This is the case we consider now.

(Stochastic differential equations with multiplicative
white noise will always be taken in the Stratonovich 
sense.)

\subsection{Random frequency}

\begin{figure}[h] 
\begin{center}
\includegraphics*[bb=120 430 520 700, width=0.5\textwidth]{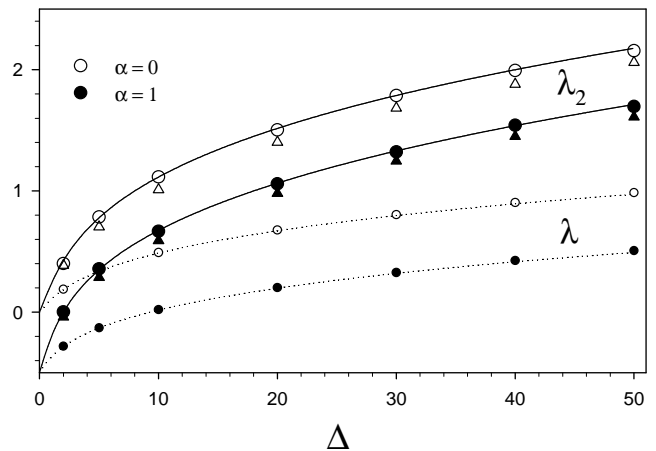}
\end{center}
\caption{
Lyapunov exponents versus noise strength for the
harmonic oscillator with random frequency.
Symbols correspond to numerical results for
$\lambda$ (small circles) and
$\lambda_2$ (large circles: importance-sampling; 
             triangles:     simple Gaussian sampling).  
Two values of the damping constant were used:
$\alpha=0$ (hollow symbols) and $\alpha=1$ (filled symbols). 
In both cases $\kappa_0=1$. 
In all cases we averaged  over $10^3$ trajectories. 
Resampling time was set to $t_{\rm res}=1.0$. 
Lines correspond to exact theoretical expressions.}
\label{fig1} 
\end{figure}

Let us first study the situation where the damping
$\alpha$ is a constant and
\begin{equation}  \label{kappa}
\kappa(t)=\kappa_0 + \xi(t)   \, ,
\end{equation}
where $\xi(t)$ is a zero-mean Gaussian white noise
with correlation function
\begin{equation}  \label{deltacorr}
\langle \xi(t) \, \xi(t') \rangle = 
\Delta \, \delta(t-t')  \, .
\end{equation} 
With these definitions one has
\begin{equation}  \label{B_RF} 
{\bf B}=
\left(
\begin{array}{ccc}
         0 &  0       &  2            \cr
         0 & -2\alpha & -2 \kappa_0   \cr
 -\kappa_0 &  1       &   -\alpha
\end{array}
\right)
+
\xi(t)
\left(
\begin{array}{ccc}
   0 &  0 & 0  \cr
   0 &  0 & 2  \cr
   1 &  0 & 0 
\end{array}
\right) \, .
\end{equation}
After substitution into Eq.~(\ref{cum}) we readily obtain
\begin{equation} \label{K_RF}
{\bf K}=
\left(
\begin{array}{ccc}
         0 &  0       &  2            \cr
    \Delta & -2\alpha & -2 \kappa_0   \cr
 -\kappa_0 &  1       &   -\alpha
\end{array}
\right) \, .
\end{equation}
The generalized exponent $\lambda_2$ can now be calculated
from Eq.~(\ref{lambdastar}).
A closed expression for the standard Lyapunov exponent was
derived by Mallick and Peyneau \cite{mallick}.
As an example, we display in Fig.~\ref{fig1} analytical and
numerical results for both exponents.
Given that the theoretical expressions are exact, 
this comparison constitutes a rigorous test for the 
numerical methods.
We see that, even for relatively small samples, 
the importance-sampling calculation agrees 
perfectly with the theory. 
The Gaussian sampling, though not perfect, provides
a reasonably good approximation.

Clearly both exponents, $\lambda_2$ and $\lambda$, do 
not coincide. 
This is to be expected whenever fluctuations in the 
frequency/damping are large as compared to their average 
values \cite{schomerus02,zillmer03}. 

The higher order exponents $\lambda_{2J}$, with $J=2,3,\ldots$
are obtained by diagonalizing matrices of size $2J-1$.
Such matrices describe the evolution of the moments 
$\langle q^n p^m\rangle$, 
with $m+n=2J$, and have simple analytical expressions 
\cite{schomerus02,zillmer03}.
An example involving a higher order exponent will be shown
in Sec.~\ref{sec6}

\subsection{Random damping}

Now we consider a harmonic oscillator with constant 
frequency but in an environment with fluctuating damping 
coefficient   
\begin{equation}  
\alpha(t)=\alpha_0 + \xi(t)   \, ,
\end{equation}
where $\xi(t)$ is again zero-mean Gaussian white noise, 
with correlation given by Eq.~(\ref{deltacorr}). 
\begin{figure}[hb!]
\begin{center}
\includegraphics*[bb=110 170 520 660, width=0.5\textwidth]{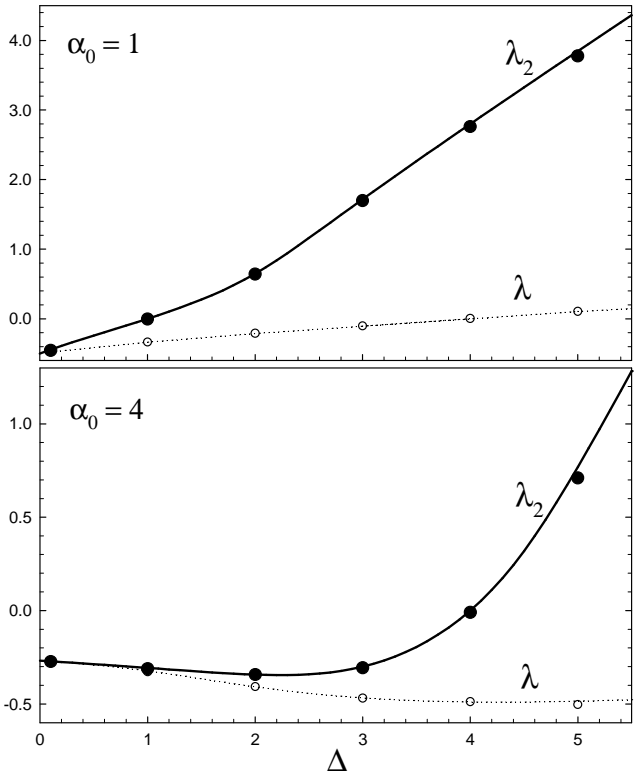}
\end{center}
\caption{Lyapunov exponents vs. noise strength for the harmonic 
oscillator with random damping.
Lines represent exact theoretical results.
Symbols correspond to numerical calculations for
$\lambda_2$ (large circles) and 
$\lambda$   (small circles).  
Importance-sampling Monte Carlo was used in the case of $\lambda_2$.
We chose two values for the average damping coefficient: 
$\alpha_0=1$ (top panel), $\alpha_0=4$ (bottom panel). 
In both cases $\kappa=1$. 
In all cases we averaged  over $10^3$ trajectories. 
Resampling time was set to $t_{\rm res}=0.2$. }
\label{fig2}
\end{figure}

Now the matrix $\mathbf{B}$ is decomposed as
\begin{equation} 
{\bf B}=
\left(
\begin{array}{ccc}
         0 &  0       &  2            \cr
         0 & -2\alpha_0 & -2 \kappa   \cr
 -\kappa &  1       &   -\alpha_0
\end{array}
\right)
+
\xi(t)
\left(
\begin{array}{ccc}
   0 &  0 & 0  \cr
   0 &  2 & 0  \cr
   0 &  0 & 1 
\end{array}
\right) \, .
\end{equation}
Hence, substitution into (\ref{cum}) yields 
\begin{equation} \label{K_RD}
{\bf K}=
\left(
\begin{array}{ccc}
         0 &  0                  &  2            \cr
    \Delta & -2\alpha_0 +2\Delta & -2 \kappa   \cr
   -\kappa &  1                  &   -\alpha_0 +\Delta/2
\end{array}
\right) \, .
\end{equation}
Upon diagonalizing $\mathbf{K}$ we obtain $\lambda_2$.
Figure~\ref{fig2} presents the comparison of numerical 
and analytical results for $\lambda$ and $\lambda_2$ as 
the noise intensity is varied (exact theoretical results 
for $\lambda$ were extracted from Ref.~\cite{leprovost}).

Concerning the numerical calculation of $\lambda_2$,
besides noting the excellent agreement with the theory, 
it must be said that the importance-sampling method behaved 
in a very robust way
both for random frequency and random damping.
Changing sample size, simulation time, and resampling time 
$t_{\rm res}$ 
\cite{vanneste10} within reasonable limits did not 
appreciably affect the result for $\lambda_2$.
However, if one increases $t_{\rm res}$ beyond
certain bounds, then the method becomes inefficient,
as very large samples are necessary to guarantee
convergence to the correct results.

\section{Correlated noise}
\label{sec5}

For white noise fluctuations, either in the frequency or in 
the damping, we have verified in the previous section
that the theory for $\lambda_2$ is in agreement with numerical 
results, provided the latter are obtained using importance
sampling. 

Now we shall analyze the effect of introducing noise 
correlations. 
We consider the case of a random frequency, 
as in Eq.~(\ref{kappa}), 
but now the noise is a zero-mean Ornstein-Ulhenbeck 
process, i.e., with correlation function
\begin{equation}
\label{corrOU}
\langle\xi(t)\xi(t')\rangle=
\frac{\Delta}{2\tau}\exp(-|t-t'|/\tau) \equiv
\sigma^2 \exp(-|t-t'|/\tau) \, .
\end{equation}
For simplicity we set $\alpha=0$ and $\kappa_0=0$. 
By inserting (\ref{B_RF}) into (\ref{cum}), the 
second-cumulant matrix becomes 
\begin{equation}  \label{K_2}
\mathbf{K}^{(2)}=\left(
\begin{array}{ccc}
               0 &                  0 &  2                 \cr
          \Delta & -2\Delta \, \tau^2 &  0                 \cr
  \Delta \, \tau &                  1 & -2 \Delta \, \tau^2 
\end{array}
\right) \,.
\end{equation}
Notice that in the limit $\tau \to 0$ the white-noise case 
is recovered. 

\begin{figure}[h] 
\begin{center}
\includegraphics*[bb=100 420 520 690, width=0.5\textwidth]{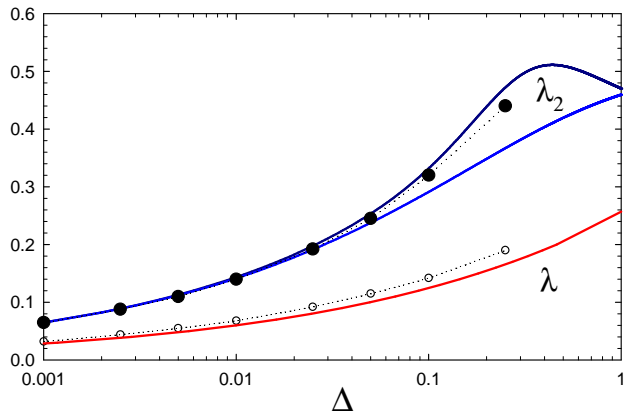}
\end{center}
\caption{\label{fig3}  
(Color online.) 
Harmonic oscillator with correlated random frequency.
Symbols indicate numerical results for
$\lambda_2$ (full circles) and 
$\lambda$ (hollow circles) 
as a function of the noise amplitude $\Delta$ 
(averages over $10^4$ trajectories). 
Parameters are $\alpha=0$, $\kappa_0=0$ and $\tau=1$. 
Resampling time was set to $t_{\rm res}=20$.
Solid lines correspond to theoretical results.
For $\lambda_2$ we used the cumulant expansion 
(blue: truncation at the second cumulant; 
dark blue: including the fourth cumulant).
An approximate analytical expression for $\lambda$ (red)
is also shown.}
\end{figure}

In the presence of correlations the second-order truncation 
of the cumulant expansion (\ref{cum}) is not exact.
In order to improve the theory one must calculate higher
cumulants. 
For the present case the third cumulant is null.
Explicit expressions for the fourth cumulant were given 
by Fox \cite{fox}, Breuer et al. \cite{breuer}, and 
Tessieri \cite{tessieri02}.
A somewhat lengthy calculation (sketched in Appendix~\ref{app1})
leads to the following result for the fourth order 
approximation to $\mathbf{K}$:
\begin{equation}  \label{K_4}
\mathbf{K}^{(4)}=\mathbf{K}^{(2)}+ \frac{1}{2}\Delta^2\tau^3
\left(
\begin{array}{ccc}
     0 &         0 &         0  \cr
    13 &  74\tau^2 &   -57\tau  \cr
17\tau & 173\tau^3 & -99\tau^2 
\end{array}
\right) \,.
\end{equation}
The comparison between numerical and theoretical results 
for $\lambda_2$ is presented in Fig.~\ref{fig3}. 
For completeness we also show numerical calculations
of the standard Lyapunov exponent, together with 
an approximate theoretical expression obtained along the 
lines of Ref.~\cite{mallick} (see Appendix~\ref{app2}).

We see in Fig.~\ref{fig3} that the inclusion of the fourth 
cumulant contribution noticeably extends the domain of validity 
of the theory into the region of larger noise amplitudes
(with respect to the second-order approximation).
Higher cumulants can also be calculated, but the required
effort quickly becomes unbearable.   
For instance, the sixth cumulant demands the calculation of
more than 100 terms (see Appendix~\ref{app1}).
Anyway, the theory being perturbative, by increasing the 
amplitude of the noise and/or the correlation time, one eventually
arrives at a point were the cumulant expansion completely 
breaks down. 
  
The perturbation parameter controlling the convergence of the
cumulant expansion is the so-called Kubo number $\varepsilon$. 
General considerations led van Kampen \cite{vankampen} to
conclude that the Kubo number is the product of the amplitude
of the fluctuations and the correlation time, that is
$\sigma \tau$.
However, in the present case it is clear that such a 
combination is not adimensional. 
The correct Kubo number is instead 
\begin{equation}
\varepsilon = 
\sigma \tau^2  = 
\sqrt{\frac{\Delta \tau^3}{2}} \, . 
\end{equation}
This can be checked explicitly from the second and fourth
cumulants above.
Consider, for instance, the element ${\bf K}_{21}$, which 
dominates the Lyapunov exponent for small correlation
times:
\begin{equation}
{\bf K}_{21} = 
\Delta + 
\frac{13}{2}\Delta^2\tau^3 + \dots =
\Delta \left(1+ \frac{13}{2}\Delta\tau^3 + \dots \right) \, . 
\end{equation}
In the white-noise limit, i.e., $\tau \to 0$ with $\Delta$
fixed, the Kubo number tends to zero --as it should be.

On the numerical side, we comment that for large noise amplitudes 
the convergence to the limiting values 
is much slower than in the white-noise cases.
The points in Fig.~\ref{fig3} were obtained by a double 
limiting procedure. 
For a fixed resampling time $t_{\rm res}$, we increased the
number of samples until convergence was reached. 
Then we iterated the scheme for increasing values of 
$t_{\rm res}$ until a stable value for $\lambda_2$ was obtained. 
The larger the resampling time, the larger the number of
samples to keep the error within the chosen bounds.

\section{Poisson shot noise}
\label{sec6}

In a dilute gas with short-range interactions,
phase-space coordinates evolve trivially in-between 
collisions. 
During collisions, positions remain essentially unchanged
while momenta experience sudden jumps.
The same description applies to tangent-space coordinates.
Thus, in a mean-field setting, the tangent dynamics of
a representative (effective) particle is described by
Eq.~(\ref{tangent2}), the stochastic frequency corresponding
to Poisson shot noise: 
\cite{hanggi80a,hanggi80b,hanggi81,lindenberg80,west80,
vankampen80,vandenbroeck82,lindenberg}
\be
\kappa(t)=\sum_i A_i \, \delta(t-t_i)  \, .
\label{poisson}
\ee
Neglecting correlations among collisions the amplitudes
$A_i$ will be modeled by independent stochastic variables
(identically distributed). 
Accordingly, the succession of collision times $\{t_i\}$
constitutes a Poisson process.

The random oscillator (\ref{tangent2}) with Poisson frequency
(\ref{poisson}) was solved by van Kampen \cite{vankampen80}
(including damping and additive noise).
He derived an exact integro-differential equation for the 
probability distribution $P(q,p,t)$ from where the evolution 
of the moments $\left \langle q^n p^m \right \rangle$ can be 
sistematically obtained \cite{vankampen80}.
For the second moments one gets
\begin{equation}  
\frac{d}{dt}
\left(
\begin{array}{c}
\langle q^2 \rangle \cr
\langle p^2 \rangle \cr
\langle qp  \rangle
\end{array}
\right)  
=
\left(
\begin{array}{ccc}
               0           & 0 &  2                         \cr
  \rho \langle A^2 \rangle & 0 & -2 \rho \langle A \rangle  \cr
 -\rho \langle A   \rangle & 1 &  0  
\end{array}
\right)
\left(
\begin{array}{c}
\langle q^2 \rangle \cr
\langle p^2 \rangle \cr
\langle qp  \rangle
\end{array}
\right)           \, ,
\end{equation}
where $\rho$ is the collision frequency. 

Remarkably the expression above can be shown to coincide 
with the result of the second-order cumulant approach 
(\ref{eqK},\ref{cum}).
However, the higher-order cumulants of $\kappa(t)$ are not 
null, rather, they are delta-correlated  
\cite{lindenberg80,west80,vankampen80}.
Just they do not affect the asymptotic growth of the second 
moments.

The equations for the fourth moments 
$\langle q^4     \rangle$,
$\langle p^4     \rangle$,
$\langle q^2 p^2 \rangle$,
$\langle q^3 p   \rangle$,
$\langle q   p^3 \rangle$ 
can also be calculated without much effort. 
The corresponding matrix reads
\begin{equation}  
\left(
\begin{array}{ccccc}
 0 & 0 & 0 & 4 & 0 \cr
   \rho \langle A^4 \rangle & 
                0           &  
 6 \rho \langle A^2 \rangle &  
-4 \rho \langle A^3 \rangle &
-4 \rho \langle A   \rangle    \cr
   \rho \langle A^2 \rangle & 
                0           &  
                0           &  
-2 \rho \langle A   \rangle &
                2              \cr
-  \rho \langle A   \rangle & 
                0           &  
                3           &  
                0           &
                0              \cr
-  \rho \langle A^3 \rangle & 
                1           &  
-3 \rho \langle A   \rangle &  
 3 \rho \langle A^2 \rangle &
                0   
\end{array}
\right) \, ,
\end{equation}
from where one extracts the fourth-order generalized exponent 
$\lambda_4$. 

\begin{figure}[h] 
\begin{center}
\includegraphics*[bb=90 420 520 690, width=0.5\textwidth]{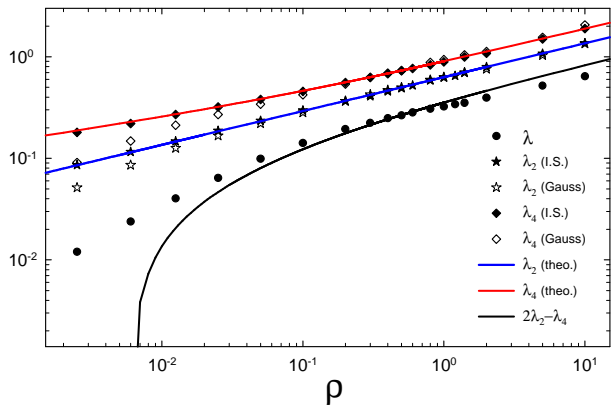}
\end{center}
\caption{   
(Color online.) Harmonic oscillator with Poisson-shot-noise frequency. 
We show the Lyapunov exponents $\lambda$, $\lambda_2$ and $\lambda_4$ 
as a function of collision frequency $\rho$.
Red/blue lines indicate theoretical estimates for 
$\lambda_4$/$\lambda_2$, and the corresponding symbols 
stand for numerical calculations using either simple-Gaussian 
sampling (open symbols) or importance-sampling (full symbols).
Shown is also the theoretical result for $2 \lambda_2 - \lambda_4$
(black line), which is an estimate of the standard Lyapunov 
exponent $\lambda$ (circles, numerical).}
\label{fig4}
\end{figure}

Figure~\ref{fig4} shows that, when importance-sampling is used,
the agreement between theory and numerics is excellent.
On the other side, simple sampling (plus a Gaussian approximation)
leads to deviations from the theory, which become stronger as 
collision frequency (``density'') is lowered.
Of course this disagreement is a consequence of the nonGaussianity 
of the distribution of finite-time Lyapunov exponents,
and can also be observed when comparing 
$\lambda$ {\em vs} $2 \lambda_2 - \lambda_4$ 
(this can be thought of as a failure of the replica trick 
\cite{deoliveira96,castiglione} in its crudest version).

The numerical method worked satisfactorily, the 
relation between parameter values and efficiency
being similar to the white-noise cases analyzed 
in Sect.~\ref{sec4}.

\section{Final remarks}
\label{sec7}

We analysed the random harmonic oscillator as a simplified
model of the tangent dynamics of many-particle systems.
In spite of its relative simplicity, this model already 
exhibits some of the essential features and characteristics
of high-dimensional systems.

Specifically, we were able to assess the performance of the
importance-sampling approach for the numerical calculation
of generalized Lyapunov exponents. 
In all the considered cases --some of which unaccessible by
standard sampling methods-- we confirmed that the method works 
satisfactorily, and developed some intuition about the 
appropriate values of the parameters (i.e., resampling time
and number of samples) that result in a faster convergence. 

On the theoretical side, we carried out several tests of the 
cumulant approach in nontrivial cases, i.e., for frequencies 
corresponding to Ornstein-Uhlenbeck and Poisson processes. 
In particular, we identifyed the correct perturbative parameter 
(Kubo number) and --not unsurprisingly-- verified that the
second-order truncation of the cumulant series gives the exact 
second-order generalized exponent $\lambda_2$ for the case of 
Poisson shot 
noise.  

Concerning the application of the cumulant approach to dilute
gases, we note that in this case the tangent dynamics can be 
thought to be driven by multivariate Poisson noise.
Accordingly the second-order 
truncation could indeed produce the exact $\lambda_2$ --like in 
the one-dimensional problem.
However, the verification of this expectation would require the
numerical calculation of $\lambda_2$ for a Hamiltonian, i.e.,
nonstochastic, system.
In order to implement an importance-sampling algorithm for this
case one should somehow introduce noise in the dynamics,
then calculate $\lambda_2$ as a function of the noise intensity,
and extrapolate the results to zero noise \cite{giardina06}.
Several ideas for constructing such an algorithm are currently under 
investigation.

\section*{Acknowledgements:}

We acknowledge Brazilian agencies Faperj and CNPq for partial 
financial support. \\[5mm]

\appendix

\section{Fourth cumulant} 
\label{app1}

Here we briefly describe the calculation of the 
fourth-cumulant contribution to the generalized 
Lyapunov exponent of the Ornstein-Uhlenbeck oscillator,
i.e., the rightmost term in Eq.~(\ref{K_4}). 
In general, this contribution reads: \cite{fox,breuer,tessieri02}
\be
K_4(t) \equiv 
\mathbf{K}^{(4)}-\mathbf{K}^{(2)} = e^{B_0 t} Q_4(t) \, e^{-B_0 t} \, ,
\ee
where 
\bea
Q_4(t) & = & \int_0^t dt_1 \int_0^{t_1} dt_2 \int_0^{t_2} dt_3 
\nonumber \\
& & 
\left( 
\;
\left \langle 
\tilde{B_1}(t) \tilde{B_1}(t_1) \tilde{B_1}(t_2) \tilde{B_1}(t_3)
\right \rangle 
\right .
\nonumber \\
& &
-\left \langle \tilde{B_1}(t)   \tilde{B_1}(t_1) \right \rangle 
 \left \langle \tilde{B_1}(t_2) \tilde{B_1}(t_3) \right \rangle
\nonumber \\
& & 
-\left \langle \tilde{B_1}(t)   \tilde{B_1}(t_2) \right \rangle 
 \left \langle \tilde{B_1}(t_1) \tilde{B_1}(t_3) \right \rangle 
 \nonumber \\
& &
\left .
-\left \langle \tilde{B_1}(t)   \tilde{B_1}(t_3) \right \rangle 
 \left \langle \tilde{B_1}(t_1) \tilde{B_1}(t_2) \right \rangle  
 \right)
 \, ,
\eea
with
\be
\tilde{B_1}(t)= e^{-B_0 t} B_1(t) \, e^{B_0 t} \, .
\ee
For the large times we are interested in, i.e., $t \gg \tau$, 
$K_4(t)$ becomes time independent. 
Next we note that matrices $\tilde{B_1}$ are proportional
to the scalar Ornstein-Ulhenbeck process $\xi(t)$.
So, one must only calculate two- and four-time correlators
of $\xi(t)$. 
By virtue of the Gaussian property \cite{foxPR78}, the four-time 
correlator is expressible as a sum of products of two-time 
functions (\ref{corrOU}).
Finally one calculates the triple integrals and takes the 
limit $t \to \infty$ (with the help of an appropriate
software, e.g., Mathematica \cite{wolfram}), arriving
thus at the desired result (\ref{K_4}).

\vspace{1pc}
\section{Lyapunov exponent} 
\label{app2}

Here we sketch the steps leading to the approximate expression 
for the Lyapunov exponent (of the random-frequency 
Ornstein-Ulhenbeck oscillator) 
that is plotted in Fig.~\ref{fig3}.
We have simply adapted the calculations of Mallick and Peyneau 
\cite{mallick} to the case $\kappa_0=0$. 

In the absence of damping, the Lyapunov exponent can be 
obtained as \cite{mallick}
\begin{eqnarray} 
\lambda 
&=& 
\lim_{t\to\infty} \frac{1}{2t} 
\langle \ln (q^2 +\dot{q}^2) \rangle    \\ 
&=& 
\lim_{t\to\infty}\frac{1}{2} \frac{d}{dt} 
\langle \ln  (q^2 +\dot{q}^2)  \rangle  \\ 
&=&   
\lim_{t\to\infty}  
(\langle y \rangle + \frac{1}{2} 
\frac{d}{dt}\langle \ln(y^2+1) \rangle)  \\ 
&=&    
\lim_{t\to\infty}  \langle y \rangle \, ,
\end{eqnarray}
where $y=\dot{q}/q$.
From Eq.~(\ref{tangent2}) one sees that $y$ obeys the following 
nonlinear equation
\begin{equation}
\dot{y}=-y^2
+\eta(t)  \, . 
\end{equation} 
We will first find the exact expression for $\lambda$ 
when the noise is white (intensity $\Delta$).
In this case, the associated 
Fokker-Planck equation for $P(y,t)$, i.e.,
\be
\partial_t P = \partial_y(y^2P)+\frac{\Delta}{2}\partial_{yy} P \, , 
\ee
has the  following steady state solution:
\begin{equation}
P_{\rm ss}(y)= 
N \, e^{-2 y^3/(3\Delta)} 
\int_{-\infty}^y e^{2 x^3/(3 \Delta)} dx \,,
\end{equation}
where $N$ is a normalization constant. 
By averaging over the steady state we obtain 
\begin{equation}  
\label{laW}
\lambda^{\rm w}(\Delta) = 
\int_{-\infty}^\infty dy \, y \, P_{\rm ss}(y) =
\frac{\sqrt{\pi}}{\Gamma(\frac{1}{6})}
\left(\frac{3\Delta}{4}\right)^\frac{1}{3} 
\simeq 0.2893 \;\Delta^{1/3} \, .
\end{equation}

In the case of an arbitrary correlation time $\tau$, 
by using a mean-field approximation 
(``decoupling ansatz'' \cite{mallick}), one can derive 
the following equation for $\lambda$: 
\begin{equation}  
\lambda(\Delta,\tau) \simeq
\lambda^{\rm w} \left(\frac{\Delta}{1+2\tau \lambda(\Delta,\tau)}\right)\,.
\end{equation} 
So, the final result comes in the form of an implicit equation:
\begin{equation}
\lambda(\Delta,\tau) \simeq 0.289 
\left( \frac{\Delta}{1+2\tau\lambda(\Delta,\tau)} \right)^\frac{1}{3} \, .
\end{equation}
This approximate relation slightly underestimates the numerical 
results of Fig.~\ref{fig3}.

\end{document}